\documentclass[useAMS,usenatbib,onecolumn]{mn2e}
\usepackage{graphicx}
\usepackage{subfigure}

\newcommand{\mpt}{\mathrm{.}}
\newcommand{\mcm}{\mathrm{,}}

\newcommand{\Vec}[1]{ \mbox{\boldmath$ #1 $} }

\title[A genetic algorithm for the non-parametric inversion of strong
lensing systems] {A genetic algorithm for the non-parametric inversion
of strong lensing systems} \author[J. Liesenborgs, S. De Rijcke and
H. Dejonghe] {J. Liesenborgs$^1$\thanks{Corresponding author:
jori.liesenborgs@uhasselt.be}, S. De Rijcke$^2$\thanks{Postdoctoral
Fellow of the Fund for Scientific Research - Flanders
(Belgium)(F.W.O)} and H. Dejonghe$^2$\\ $^1$ Expertisecentrum voor
Digitale Media, Universiteit Hasselt, Wetenschapspark 2, B-3590,
Diepenbeek, Belgium \\ $^2$ Sterrenkundig Observatorium, Universiteit
Gent, Krijgslaan 281, S9, B-9000, Gent, Belgium}

\begin{document}
	
\date{} 
	
\pagerange{\pageref{firstpage}--\pageref{lastpage}} \pubyear{2005}
	
\maketitle \label{firstpage} 

\begin{abstract} 
We present a non-parametric technique to infer the projected-mass
distribution of a gravitational lens system with multiple
strong-lensed images. The technique involves a dynamic grid in the
lens plane on which the mass distribution of the lens is approximated
by a sum of basis functions, one per grid cell. We used the projected
mass densities of Plummer spheres as basis functions. A genetic
algorithm then determines the mass distribution of the lens by forcing
images of a single source, projected back onto the source plane, to
coincide as well as possible. Averaging several tens of solutions
removes the random fluctuations that are introduced by the
reproduction process of genomes in the genetic algorithm and
highlights those features common to all solutions. Given the positions
of the images and the redshifts of the sources and the lens, we show
that the mass of a gravitational lens can be retrieved with an
accuracy of a few percent and that, if the sources sufficiently cover
the caustics, the mass distribution of the gravitational lens can also
be reliably retrieved. A major advantage of the algorithm is that it
makes full use of the information contained in the radial images,
unlike methods that minimise the residuals of the lens equation, and
is thus able to accurately reconstruct also the inner parts of the
lens.
\end{abstract}
	
\begin{keywords}
gravitational lensing -- methods:~data analysis -- dark matter --
galaxies:~clusters:~general
\end{keywords}
	
\section{Introduction}

The deflection of light caused by a gravitational lens and the
amplifying and distorting effect thereof on the images of background
sources, provide us with a means to measure the total mass of the
lensing object. If only, of course, that it is possible to ``invert''
such a lensing system, i.e. to infer the mass distribution of the lens
given the positions and shapes of a set of lensed images of background
sources and the redshifts of the lens and the sources. The inversion
of gravitational lensing systems is interesting in its own right,
since it puts constraints on the spatial dark matter distribution of
the lensing objects and thus helps constrain dark matter physics
\citep{n04,d04}, but it may also contribute to cosmology. Good
reconstructions of lensing systems may help to constrain the density
parameter and the redshift evolution of the dark energy
\citep{y01,s04,m05}.

The use of the gravitational lensing effect to measure masses of
point-mass lenses (or ``stars'') was envisaged by \citet{r64} and
\citet{l64}. The idea of using simple parametric models for extended
lenses, such as galaxies or clusters of galaxies, can be traced back
to \citet{dr80}, who use a King model to estimate the mass
distribution of the galaxy that produces the two images of
QSO~0957+561A,~B. Since then a host of parametric inversion methods
has been developed and applied to observed strong lensing systems,
such as the ring cycle method of \citet{k89} or the maximum entropy
method of \citet{w94}. \citet{kn93} fitted a bimodal lensing
potential, constructed from two elliptical pseudo-isothermal
potentials, images of the cluster A370, obtained from the ground under
excellent seeing conditions. More elaborate parametric models for
lensing clusters associate a simple mass distribution, e.g. a
power-law of radius, to each galaxy and to the cluster as a whole. The
many parameters that define the lens model are then determined by a
$\chi^2$ fit to the observed images \citep{tkd98}. \citet{b05} suggest
using a gravitational lens as a cosmic magnifying glass to recover
structural details about distant sources. These authors assume a
parametric form for the lens mass distribution and use a genetic
algorithm to non-parametrically reconstruct the surface brightness
distribution of the source. If only a few sources are being lensed or
if the sources do not sufficiently cover the caustics, parametric
methods are clearly the preferred approach.

However, the multitude of arcs and distorted images visible in massive
clusters (e.g. \citealt{s87,lh88}), which are observed routinely now
at high spatial resolution with the Hubble Space Telescope
(e.g. \citealt{br05}), contain a wealth of information and call for
more flexible and model independent inversion methods. Using the
pixelation method \citep{sw97,a98}, the lens plane is divided into a
static grid. The mass in each grid cell and the source positions are
estimated so as to construct a solution that best reproduces the
observed image positions, subject to regularizing constraints that
ensure a smooth mass distribution for the lens that stays close to the
luminosity distribution. \citet{t00} perform a multipole-Taylor
expansion of the two-dimensional lensing potential, the coefficients
of which are determined by a $\chi^2$-fit to the observed images. In
\citet{d05}, a non-parametric inversion technique, called {\sc slap},
is presented and applied to an HST image of the cluster Abell 1689
\citep{d05b,k02} that, like the pixelation method, makes use of a grid
division of the mass distribution of the lens. This time, however, the
grid is dynamic:~it is refined iteratively where the mass density is
large. Recently, \citet{d05c} extended {\sc slap} to {\sc wslap} in
order to also take into account information in the weak lensing
regime. Thus, {\sc slap} and {\sc wslap} are fast and versatile tools
for inverting observed lens systems. However, both methods assume the
background galaxies to be point sources, which may lead to an
over-estimation of the central mass density of the lens in order to
focus the images into very compact sources and to physically
implausible regions with negative mass density in the lens plane.
Adding weak-lensing information alleviates the dependence of the
solution on this minimization threshold. {\sc wslap} can make use of
quadratic programming to avoid unphysical negative mass densities for
the lens. However, this limits the analysis to observables that are
linear functions of the lens mass density, such as image positions,
and does not allow to incorporate e.g. surface brightness information,
which depends non-linearly on the lens mass density. \citet{bra05}
also used weak and strong lensing data to invert the X-ray cluster
RX~J1347.5$-$1145. Their method evaluates the gravitational potential
on a non-dynamic grid. The best fitting gravitational potential is
then constructed non-parametrically by minimising a $\chi^2$ function,
starting from a  parametric priorsolution.

The ideal non-parametric lens inversion algorithm {\em (i)} should be
free of any assumptions regarding the mass distribution of the lens or
the luminosity distributions of the sources, {\em (ii)} should not
depend on any prior on the lens mass distribution or any
regularisation scheme that could bias the solution, {\em (iii)} should
not produce unphysical, i.e. negative, mass densities {\em (iv)}
should be free of any uncontrollable parameters, {\em (v)} should be
easily extendible to any kind of data, both in the strong and weak
lensing regimes, without having to change the inner workings of the
algorithm or having to worry about features like continuity or
differentiability of the objective function that is extremised. These
conceptual issues are the main motivation for this paper, rather than
computational speed. 

Genetic algorithms do an excellent job at fulfilling all these
constraints. In this paper, we describe and test a new non-parametric
lens inversion technique. The technique makes use of a dynamic grid on
which the mass distribution of the lens is approximated by a weighted
sum of basis functions and of a genetic algorithm to determine the
unknown weights. In the strong lensing regime, where multiple images
of each source are available, the following data are offered to the
algorithm:~the redshifts of the sources and the lens, and the
observed positions of the images. The genetic algorithm generates
solutions that satisfy only one minimal constraint:~the
back-projected images of a single source should overlap as well as
possible in the source plane.

We briefly describe the relevant background to gravitational lens
systems and genetic algorithms in Section \ref{sec:back}. The details
of the inversion method are given in Sect. \ref{sec:invert}. We
discuss a number of tests to which we subjected the method in
Sect. \ref{sec:sim}. Finally, our conclusions are summarized in
Sect. \ref{sec:conc}.

\section{Background} \label{sec:back}
	
\subsection{The lens equation}
		
In the thin lens approximation, the lens equation relates viewing
directions $\Vec{\theta}$, that are defined in the lens plane, to
positions $\Vec{\beta}$ in the source plane:~ \begin{equation}
\Vec{\beta}(\Vec{\theta}) = \Vec{\theta} -
\frac{D_{ds}}{D_s}\;\Vec{\hat{\alpha}}(\Vec{\theta}) \mcm
\label{eq_lenseqn} \end{equation} with $\Vec{\hat{\alpha}}$ the
deflection angle, $D_s$ the distance between the observer and the
source, and $D_{ds}$ the distance between the lens and the source. The
gravitational bending of light rays, described by the deflection angle
$\Vec{\hat{\alpha}}$, depends on the viewing direction $\Vec{\theta}$,
the mass distribution of the lens and the distance between the lens
and the observer. Here and in the following, distances should be
interpreted as angular-diameter distances. For simplicity, we will
adopt a standard CDM cosmology, with a matter density $\Omega=1$ and a
Hubble parameter $H_0 = 70$~km~s$^{-1}$~Mpc$^{-1}$, in which the
angular-diameter distance between an observer at redshift $z_1$ and a
source at redshift $z_2$ is given by \begin{equation} D(z_1,z_2) =
\frac{2c}{H_0} \frac{1}{1+z_2} \left(\frac{1}{\sqrt{1+z_1}}-
\frac{1}{\sqrt{1+z_2}} \right) \mpt \end{equation} We will always
assume that the redshifts of the lens and of the source(s) are known
to the observer. The lens equation projects the images back onto their
respective sources in the source plane. When a gravitational lens
produces multiple images of a single source, one can use the lens
equation to find, for each image, the corresponding region in the
source plane. Since all images correspond to a single source, all
back-projected images have to coincide.
						
\subsection{The Plummer lens}
		
We first describe the gravitational lens effect caused by a Plummer
sphere \citep{p11} at a distance $D_d$ from the observer. The
projected density distribution of a Plummer sphere with total mass $M$
and angular scale-length $\theta_P$ as a function of angular distance
$\theta$ is given by \begin{equation} \Sigma(\theta) = \frac{M}{\pi
D_d^2}\frac{\theta_P^2}{(\theta^2+\theta_P^2)^2}.
\label{eq_sigma_plummer} \end{equation} This mass
distribution leads to the following lens equation: \begin{equation}
\Vec{\beta}(\Vec{\theta}) = \Vec{\theta}-\frac{D_{ds}}{D_s D_d}\frac{4
G M}{c^2}\frac{\Vec{\theta}}{\theta^2+\theta_P^2} \end{equation} if
the coordinate system in the lens plane is centered on the Plummer
sphere.
			
As a first step towards inverting a given lens system, we write the
(unknown) projected mass distribution of the lens as a sum of Plummer
mass distributions, of the form given by
eq. (\ref{eq_sigma_plummer}). We chose the Plummer mass distribution
as basis function because it is well-behaved at all radii and yields a
finite total mass. The deflection angle is then simply the sum of the
deflection angles caused by each individual Plummer distribution. For
$N$ individual Plummer lenses, this yields the following lens
equation: \begin{equation} \Vec{\beta}(\Vec{\theta}) = \Vec{\theta} -
\frac{D_{ds}}{D_s D_d} \frac{4 G}{c^2} \sum_{i = 1}^N
\frac{\Vec{\theta}-\Vec{\theta}_{s,i}}{|\Vec{\theta}-\Vec{\theta}_{s,i}|^2+\theta_{P,i}^2}
M_i \mcm \label{eq_lenseqn_mplum} \end{equation} with
$\Vec{\theta}_{s,i}$ the position of the centre of a Plummer
distribution in the lens plane, $M_i$ its mass, and $\theta_{P,i}$ its
angular scale-length.
			
A given set of $R$ points in the lens plane, $\Vec{\theta}_k,\,k = 1
\ldots R$, is related to a corresponding set of $R$ points in the
source plane by a matrix equation \citep{d05}. Indeed, let $\Theta$ be
a vector of length $2R$, containing the coordinates of the points in
the image plane, in which $x$ and $y$ components alternate. Similarly,
$B$ is a vector of length $2R$ which will contain the coordinates of
the corresponding points in the source plane. The masses $M_i$ of the
Plummer distributions that make up the mass distribution of the lens
are stored in an $N$ dimensional column vector $M$. The lens equation
can then be rewritten as
\begin{equation} B = \Theta - \gamma M \mcm \end{equation} with
$\gamma$ a $2R \times N$ matrix whose components are given by:
\begin{eqnarray} \gamma_{2k-1,l} &=& \frac{D_{ds}}{D_d D_s} \frac{4
G}{c^2}
\frac{(\Vec{\theta}_k-\Vec{\theta}_{s,l})_x}{|\Vec{\theta}_k-\Vec{\theta}_{s,l}|^2+\theta_{P,l}^2}
\nonumber \\ \gamma_{2k,l} &=& \frac{D_{ds}}{D_d D_s} \frac{4 G}{c^2}
\frac{(\Vec{\theta}_k-\Vec{\theta}_{s,l})_y}{|\Vec{\theta}_k-\Vec{\theta}_{s,l}|^2+\theta_{P,l}^2}
\mpt \end{eqnarray} The problem of inverting a gravitational lens
system is thus transformed into the problem of finding the vector $M$,
given the matrices $\Theta$ and $\gamma$.
			
\subsection{Genetic algorithms} \label{genalg}
			
With genetic algorithms, one tries to breed good solutions to a given
problem. A central concept is the genome, which is an encoded
representation of a possible solution. Usually, the genome will encode
the parameters of a specific model. For a particular genome, there has
to be some kind of measure of how adequate it fits the data. This
value is usually called the fitness of the genome. The algorithm
starts with a random set of genomes: the population.  From this
population, a new one will be created using the following procedure:
\begin{itemize} \item For each genome, the fitness is calculated.
\item A new set of genomes is created by combining and copying genomes
of the current population. Selection of genomes in this reproduction
step should favor genomes with a better fitness.
\item Finally, mutations are introduced in the new population to
ensure genetic variety.  \end{itemize} When creating the new
population, the best genome is often copied without mutations. This
approach is often referred to as elitism and ensures that the best
member of the new population will perform at least as well as the
fittest member of the old population. Thus, generation after
generation, one tries to breed increasingly better solutions to a
problem. A complete overview of genetic programming techniques can be
found in \cite{koza:book}.
			
\section{The inversion method}\label{sec:invert}
	
In the following, we discuss two key features of our inversion
method: the use of a dynamic grid in the lens plane on which the
Plummer lenses are positioned (which defines the matrix $\gamma$) and
the genetic algorithm employed to breed the best approximation to the
projected mass distribution of the lens (i.e., the vector $M$).
					
\begin{figure}
\centering
\subfigure[Step one]{\includegraphics[width=0.23\textwidth]{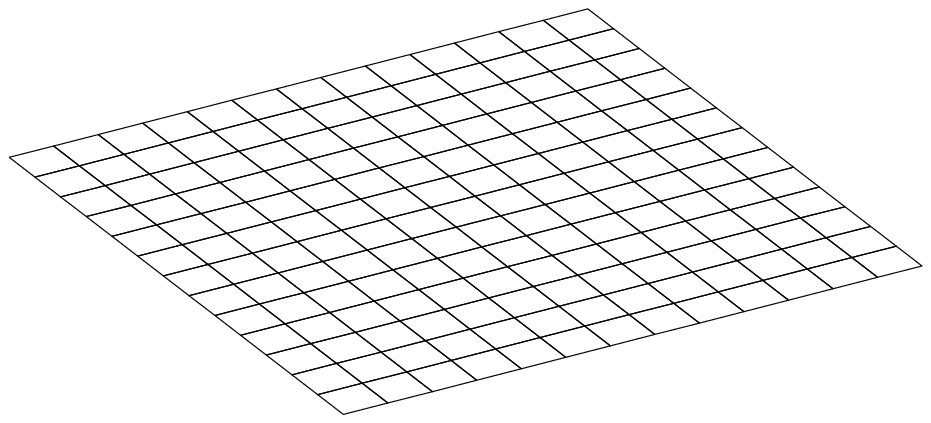}}
\subfigure[Step two]{\includegraphics[width=0.23\textwidth]{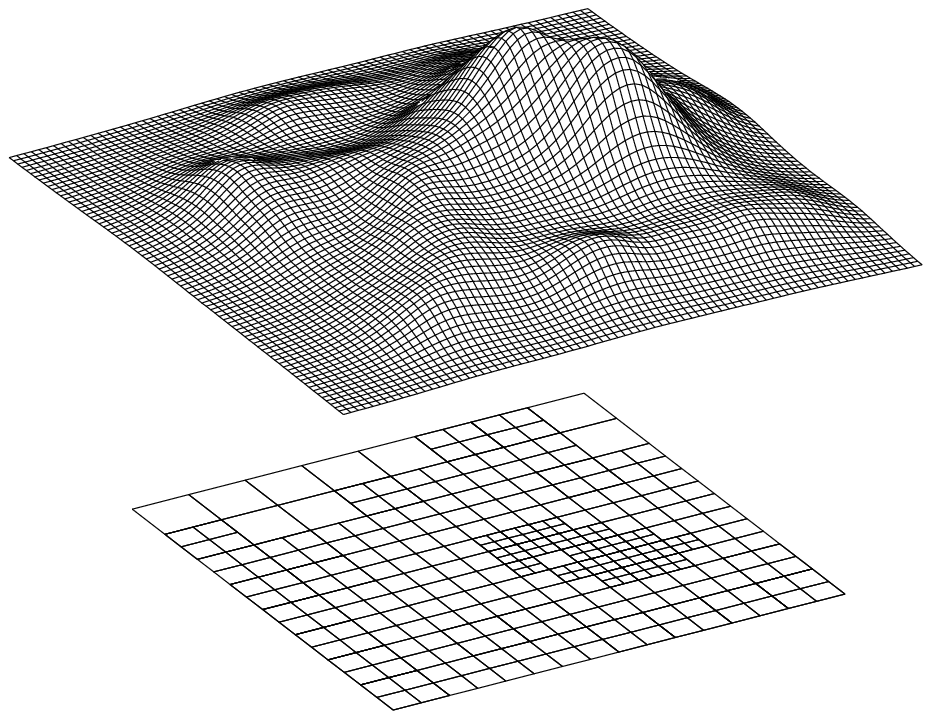}}
\subfigure[Step three]{\includegraphics[width=0.23\textwidth]{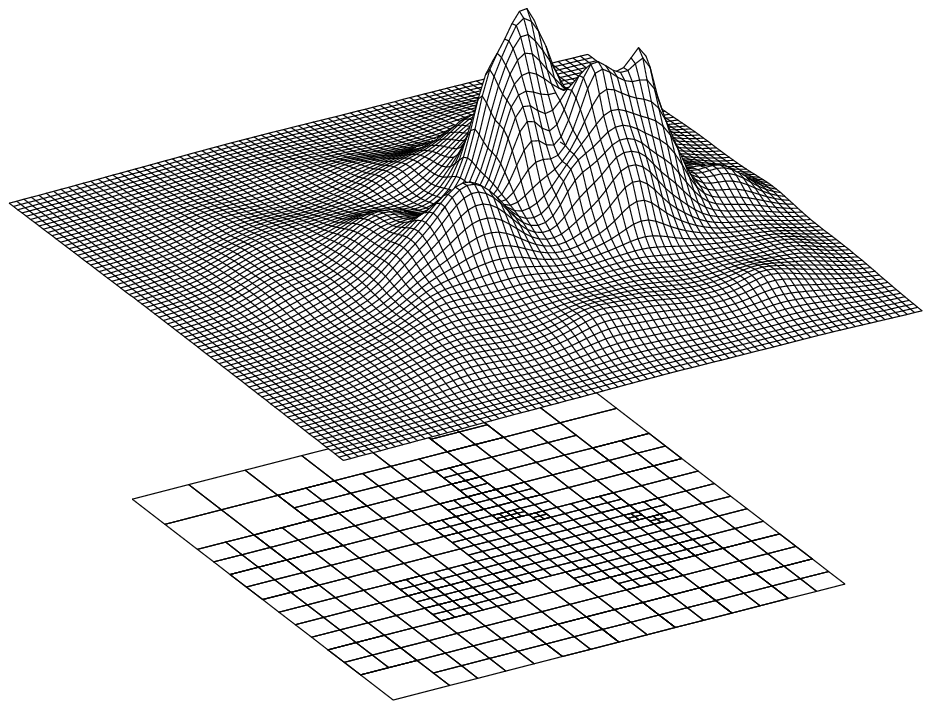}}
\subfigure[Step four]{\includegraphics[width=0.23\textwidth]{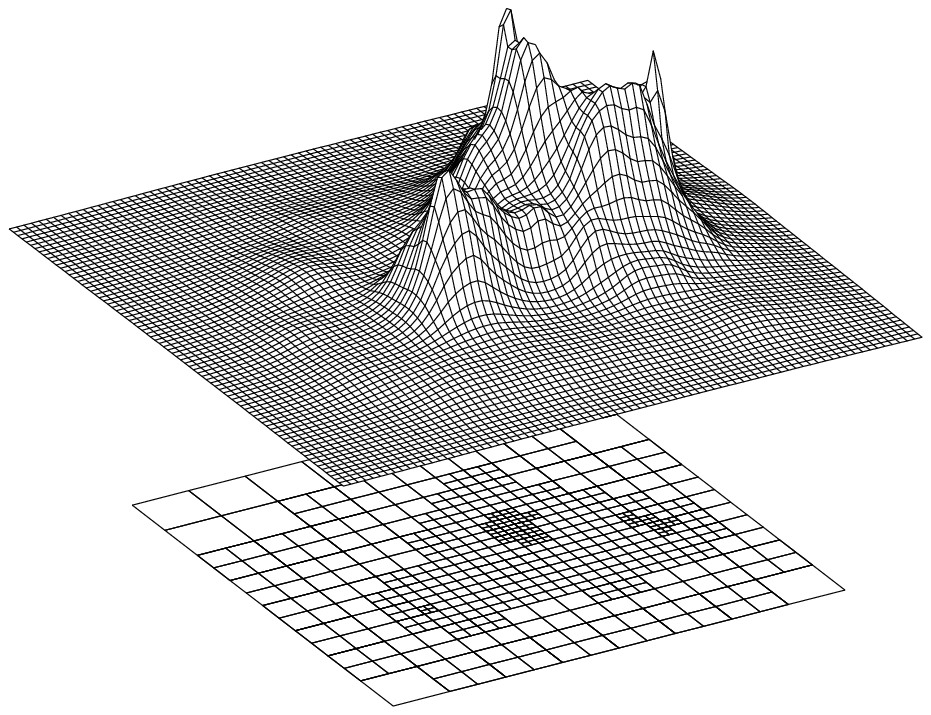}}
\addtocounter{subfigure}{-4} \caption{The use of a dynamic grid. The
grid spacing is refined in those regions where individual grid cells
contain a large fraction of the total mass of the lens or where the
projected-density gradients are large.}  \label{fig1}
\end{figure}
	
\subsection{The dynamic grid}
		
The procedure starts with a square grid, large enough to encompass the
projected mass density of the lens. At first, this area is uniformly
subdivided in square grid cells. At the centre of each grid cell, a
Plummer mass distribution is positioned. The width of each Plummer
distribution is set proportional to the side of its grid cell. We
tested which proportionality factor allows to best reproduce a wide
range of mass densities and found that a value of 1.7 yields a good
trade-off between smoothness and dynamic range. The same scale factor
was subsequently used in our lens inversion simulations. The genetic
algorithm (see subsection \ref{sub:gen}) then breeds, for this given
grid, the best solution $M$. Given this first approximation of the
total mass density, a new grid is constructed by further subdividing
grid cells that contain a large fraction of the total mass or that
reside in areas with large density gradients. This way, the new grid
will allow a better approximation of the mass density, without wasting
resources on areas which contain little mass or detail. With each cell
of this new grid a Plummer distribution is associated and the
individual masses are determined by the genetic algorithm, as before.
In our implementation, this procedure of refining the grid is repeated
unless the number of grid cells exceeds one thousand. Fig. \ref{fig1}
illustrates the procedure. At first, a uniform grid is used. With this
grid, a first estimate of the distribution is found and this is used
to create a new grid. The figure shows a few additional mass density
estimates on which new grids are based.

\subsection{The genetic algorithm} \label{sub:gen}
					
\begin{figure}
\centering
\includegraphics[width=0.48\textwidth]{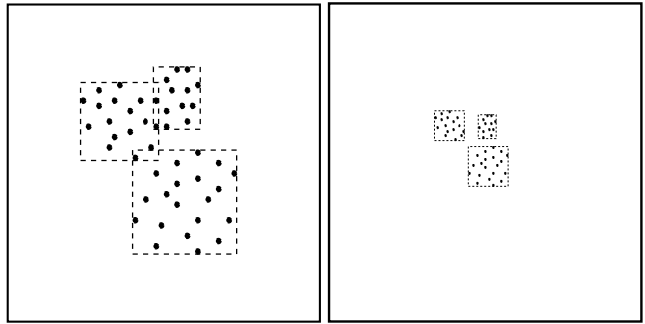}
\caption{Evaluating the fitness of back-projected images. In the left
panel, on an absolute scale, the value of the potential energy of the
imaginary springs connecting the rectangles that enclose the
back-projected images of a single source is higher than in the
situation depicted in the right panel. However, the back-projected
images in the left panel overlap, unlike those in the right panel. We
therefore scale the rectangles enclosing the back-projected images of
a single source with the mean size of the rectangles when calculating
the fitness value.}  \label{fig2}
\end{figure}
		
\subsubsection{Genome and fitness}

\begin{figure}
\centering \includegraphics[width=0.45\textwidth]{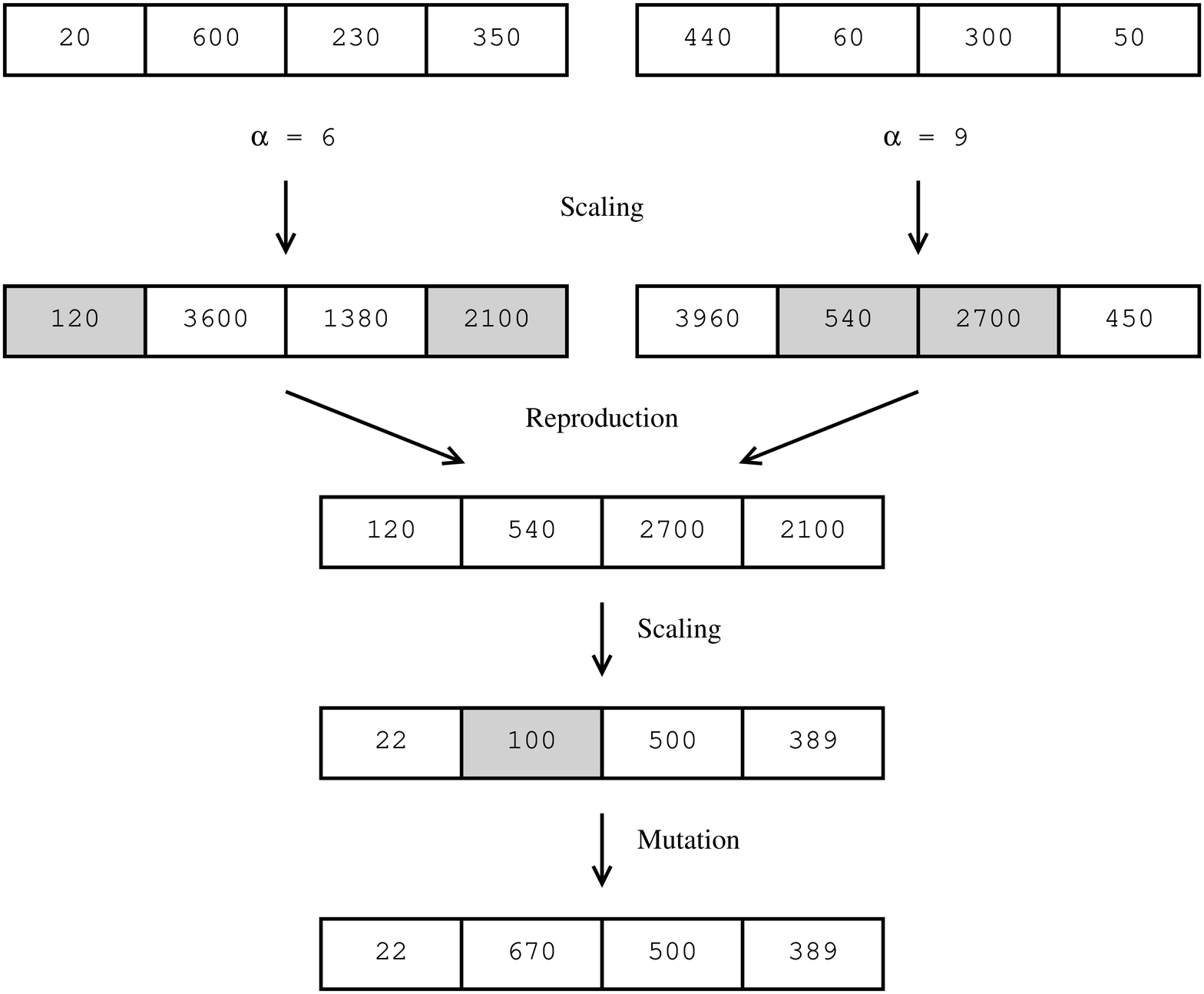}
\caption{Reproduction and mutation of genomes. See text (subsection
\ref{sub:rep}) for explanation.}  \label{fig3}
\end{figure}
						
The goal of the genetic algorithm is to determine good values for the
masses of the individual Plummer distributions which are laid out
according to a specific grid. Therefore, the genome in our genetic
algorithm will encode the masses of these Plummer distributions.
					
For a specific set of Plummer masses, we need to define a way to
evaluate how good the corresponding solution is. Since we are working
in the strong lensing regime, it is assumed that the gravitational
lens system produces multiple images of one or more sources. If one
would project these images back to the source plane using the exact
lens equation for the lens under study, one would find that
back-projected images of the same source will overlap perfectly. For
this reason, the degree in which back-projected images of the same
source overlap will be used to determine how good the suggested
solution actually is.
				
The way this is implemented is as follows. For a given solution of the
mass distribution of the lens, the images of a single source are
projected back to the source plane. The areas occupied by each image
are surrounded by rectangles: two examples are shown in Fig.
\ref{fig2}. Corresponding corners of the rectangles are connected with
imaginary springs. Consider two rectangles, each enclosing a
backprojected image. For corresponding corners, the distance is
calculated in absolute units, for example in units of arcminutes or
arcseconds. In a previous step, a length scale was calculated as the
average of the lengths of the sides of all the rectangles belonging to
a specific source. The distance between two corresponding corners is
then divided by this length, yielding a dimensionless distance
$d$. The ``potential energy'' for this pair of corners is then simply
$d^2$. Repeating this for the other three corners and adding together
the energies then gives the potential energy of these two
rectangles. For a specific source, this procedure is then done for all
pairs of backprojected images and the sum of these potential energies
is then the potential energy contribution of this source. The fitness
value of a given lens solution is the sum of the potential energies of
all sources. It is important to take into account the scaling of the
rectangles when calculating the potential energy values. Comparing the
left and right parts of Fig. \ref{fig2}, it is clear that the left
situation definitely corresponds to a better overlap, while on an
absolute scale the potential energy of the right situation will be the
lower one. For this reason, we express distances between corners of
rectangles relative to the size of the rectangles, or, in other words,
relative to the size of the source.
					
As was mentioned above, the genome represents the masses of the
individual Plummer distributions. To be more precise, the genome only
represents the relative contribution of each Plummer distribution:
each Plummer mass is represented by a dimensionless, integer number
between 0 and 1000. These numbers are stored in the vector $M$ and the
matrix product \begin{equation} \Theta' = \gamma M \end{equation} is
calculated. For the dimensionless masses to be converted into real
masses, the vector $M$ needs to be multiplied with a factor $\mu$,
bringing the lens equation in the form \begin{equation} B = \Theta -
\mu \Theta'.  \end{equation} Since $\Theta$ and $\Theta'$ are constant
column matrices, it is an easy and computationally inexpensive task to
find, for a given $M$, the factor $\mu$ that maximises the fitness,
or, in other words, for which the back-projected images of the sources
coincide best. The value of the fitness of that particular situation
is then considered to be the fitness of the genome.
				
\subsubsection{Reproduction and mutation}\label{sub:rep}
					
In our implementation, a population of 250 genomes is used. Based on
the many simulations we did (see below), 250 genomes has always led to
good solutions within an acceptable amount of time. To obtain a new
population, some genomes are copied from the original population while
others are obtained by merging two genomes. The procedure of merging
two genomes consists of a few steps which are illustrated in
Fig. \ref{fig3}.  At first, the values between 0 and 1000 of each
genome are multiplied with their best $\mu$ value to obtain the true
Plummer masses they represent. Then, for each Plummer distribution,
the procedure selects at random the mass from one of the two
genomes. Finally, these values are rescaled to integer numbers in such
a way that the largest number is 500.
									
When the new population is complete, mutations are introduced in some
genomes. In early generations, some values are simply changed to a
random number between 0 and 1000. It is for this reason that the
previous step rescaled the Plummer masses to a maximum value of
500. This way, a random change of the value will also allow a
considerable increase in mass for that Plummer distribution.

When the best fitness values of successive generations start to
converge, a new mutation rule is adopted. In this case, random integer
numbers in the interval $[-200,200]$ are generated and added to some
of the genomes' values. Resulting values which are negative or larger
than $1000$, are set to zero or $1000$ respectively. The first
mutation rule makes sure that a large range of mass densities can be
inspected. When the algorithm starts to converge near a good solution,
the second mutation rule assures that the algorithm can more closely
approach the best solution.

\subsubsection{Stopping criterion}

The algorithm can be stopped if the fitness of the best genome ceases
to improve significantly. We use the following stopping criterion : if
the fitness of the last generation, denoted by {\tt new\_fit} fulfills
the constraint $|${\tt new\_fit}$-${\tt old\_fit}$| <$ {\tt
new\_fit}/50, with {\tt old\_fit} the fitness of 150 generations ago,
the algorithm is stopped. We inspected that raising the factor of 50,
so that the code runs longer, does not significantly change the
solution, i.e. that the solution has converged. Just for safety, we
implemented an upper limit of 15000 on the number of generations but
all inversions we tried so far converged after less than 5000
generations.

\subsection{Averaging multiple solutions}
		
The genetic algorithm uses a random initial population, selects
genomes at random and introduces random mutations. Because of this,
multiple applications of the inversion procedure for a specific set of
images will in general yield slightly different solutions. These
solutions are all equally acceptable: in all cases, the back-projected
images of a single source coincide very well with each other and with
the true position of the source. Given this variety of possible
solutions, it is interesting to calculate the average of a set of
solutions. This averaging procedure will enhance the common
characteristics of all the individual lens solutions while suppressing
random fluctuations. One can also calculate the standard deviation of
these individual solutions. This will identify the regions in which
the solutions agree as well as the regions in which there is a lot of
uncertainty about the mass density. Averaging the solutions will also
increase the smoothness of the retrieved mass density.

What determines the convergence of the fitness value (see section
\ref{genalg}) is the amplitude of the mutations. Once the difference
between the best possible lens solution and the best genome becomes
comparable to or smaller than the mutation amplitude, the best genomes
of subsequent generations merely scatter around some lowest achieved
fitness value. Lowering the mutation amplitude when the fitness starts
to converge and averaging a few tens of independent solutions both
help to get as close as possible to the best possible solution.
			
Being able to create an averaged solution is an attractive feature of
our approach, but it would be of little use if the resulting mass
density would not be a good solution of the inversion problem (or a
worse solution than the individual solutions). Using simulations (see
Sect. \ref{sec:sim}), we found that the averaged solution is indeed
also a good solution, with a very high fitness, and in many cases even
does a better job than many of the individual solutions. This is
because the random mutations that occur during the reproduction
process of the genomes, cause the best solution to oscillate around
the ``true'' solution. Since averaging a set of solutions suppresses
these random fluctuations, the averaged solution can be a more faithful
realisation of the true solution than any of the individual
solutions. Also, the inversion of a gravitational lens is clearly an
ill-posed problem so it's no great supprise that multiple solutions
exist. For these reasons, the averaged solution is definitely a very
acceptable one.

\section{Simulations}\label{sec:sim}

We conducted many simulations in order to test the validity of our
approach. Having full knowledge of the original lens as well as of the
original sources, we can easily check the accuracy with which lenses
can be reconstructed in ideal circumstances, i.e. when the redshifts
of the lens and the sources are known exactly. The mass distributions
of the lenses in these simulations were created by randomly adding a
number of Plummer distributions. The number of sources, their
positions and redshifts were also chosen at random. A wide variety of
these gravitational lens systems were used to test the algorithm. In
the example that we present below, a lens with mass of the order of
$10^{15}\,M_\odot$ was positioned at $z = 0.45$ while the redshifts of
the sources were sampled from a uniform distribution in the interval
$[1.2,4.0]$.

\begin{figure*}
\subfigure{\includegraphics[origin=c,angle=-90,width=0.48\textwidth]{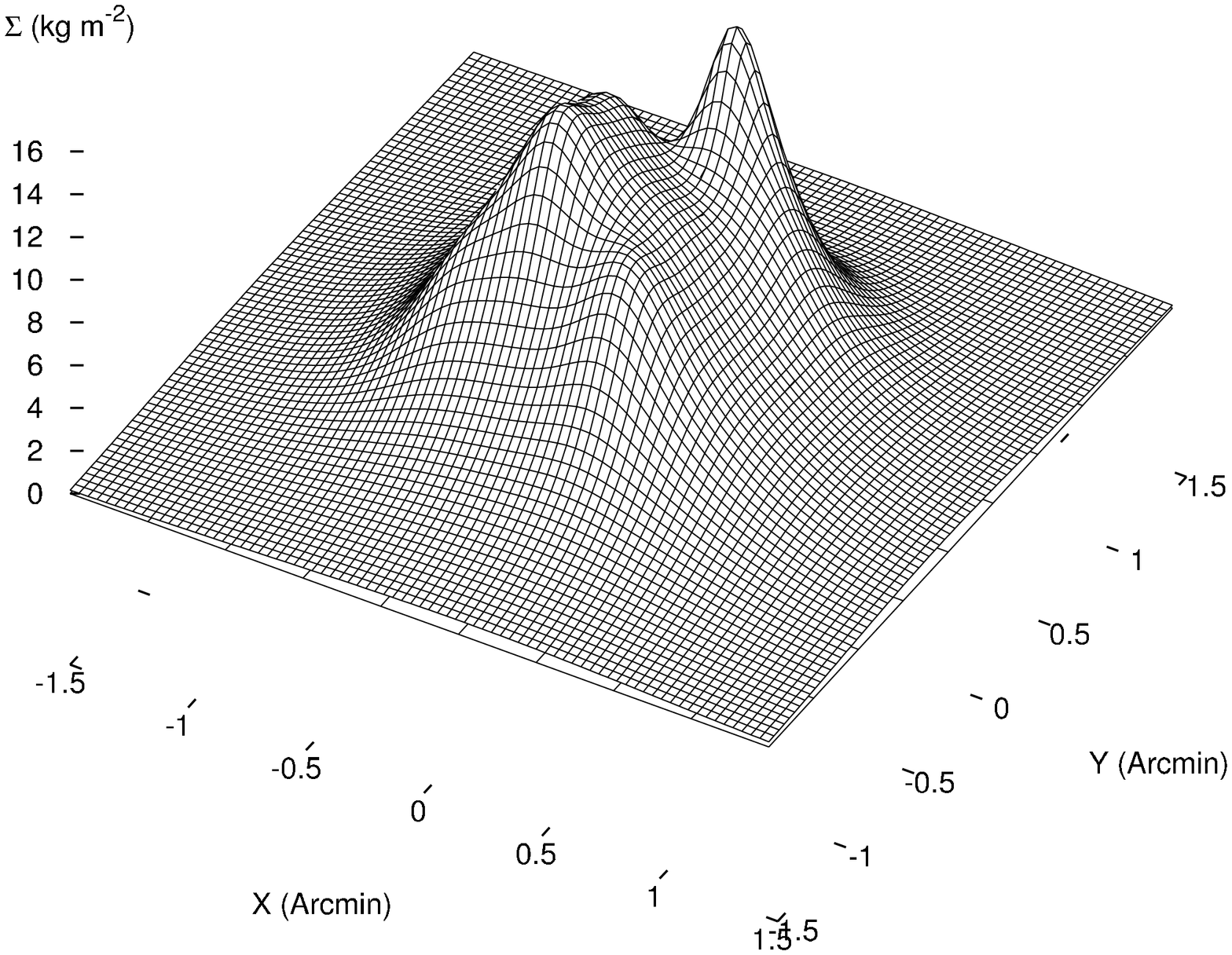}}
\subfigure{\includegraphics[width=0.48\textwidth]{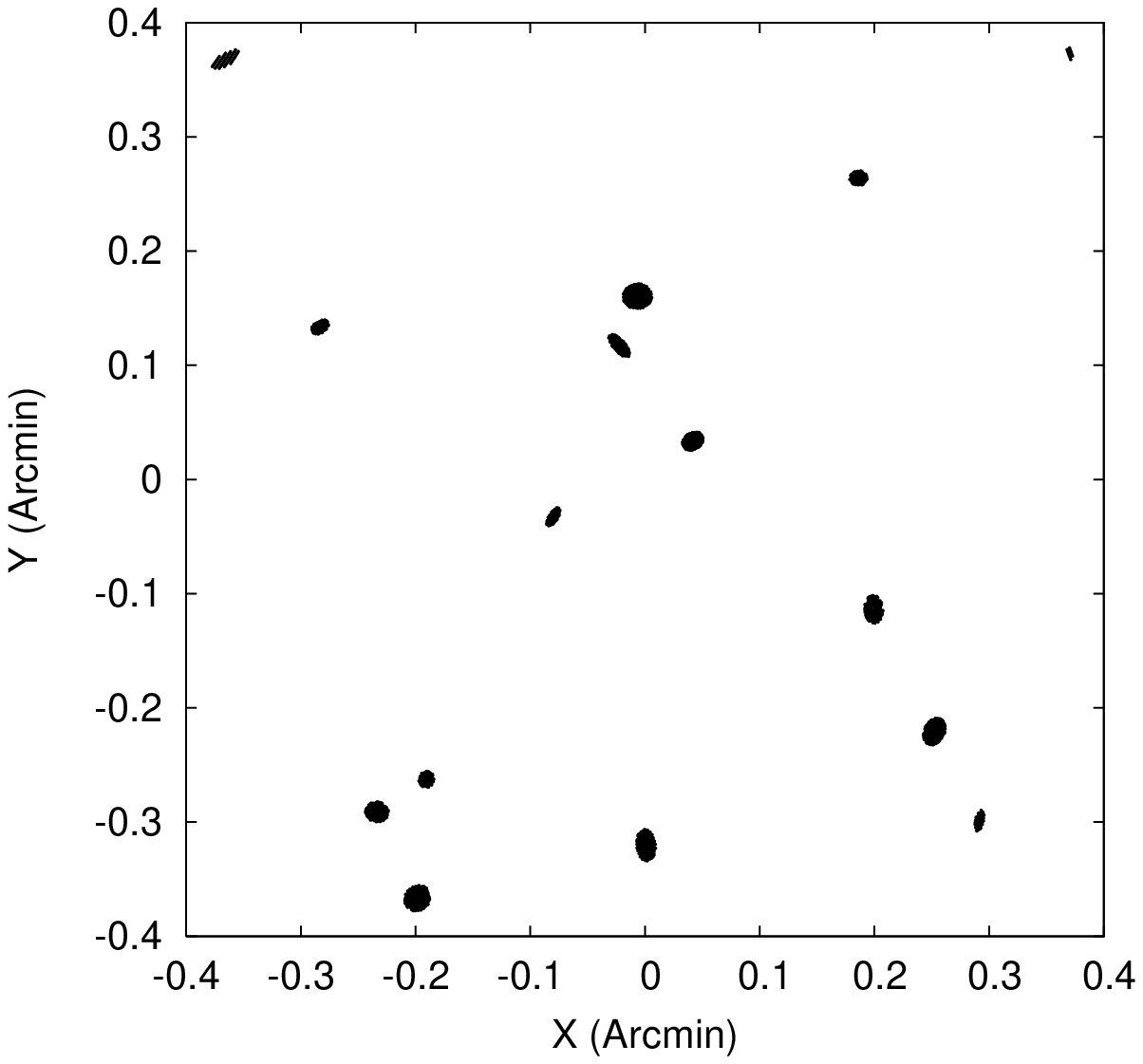}}
\addtocounter{subfigure}{-3} \caption{Left panel:~the mass
distribution of the input lens used in the simulation. The total lens
mass within a radius of $1.5${\arcmin}, which is slightly further out
than the position of the outermost image, is $0.95\times 10^{15}
\,M_\odot$. Right panel:~the positions and shapes of the 15 sources
used in the simulation within the source plane.}
\label{fig4} \end{figure*}	
	
\begin{figure*}
\subfigure{\includegraphics[width=0.48\textwidth]{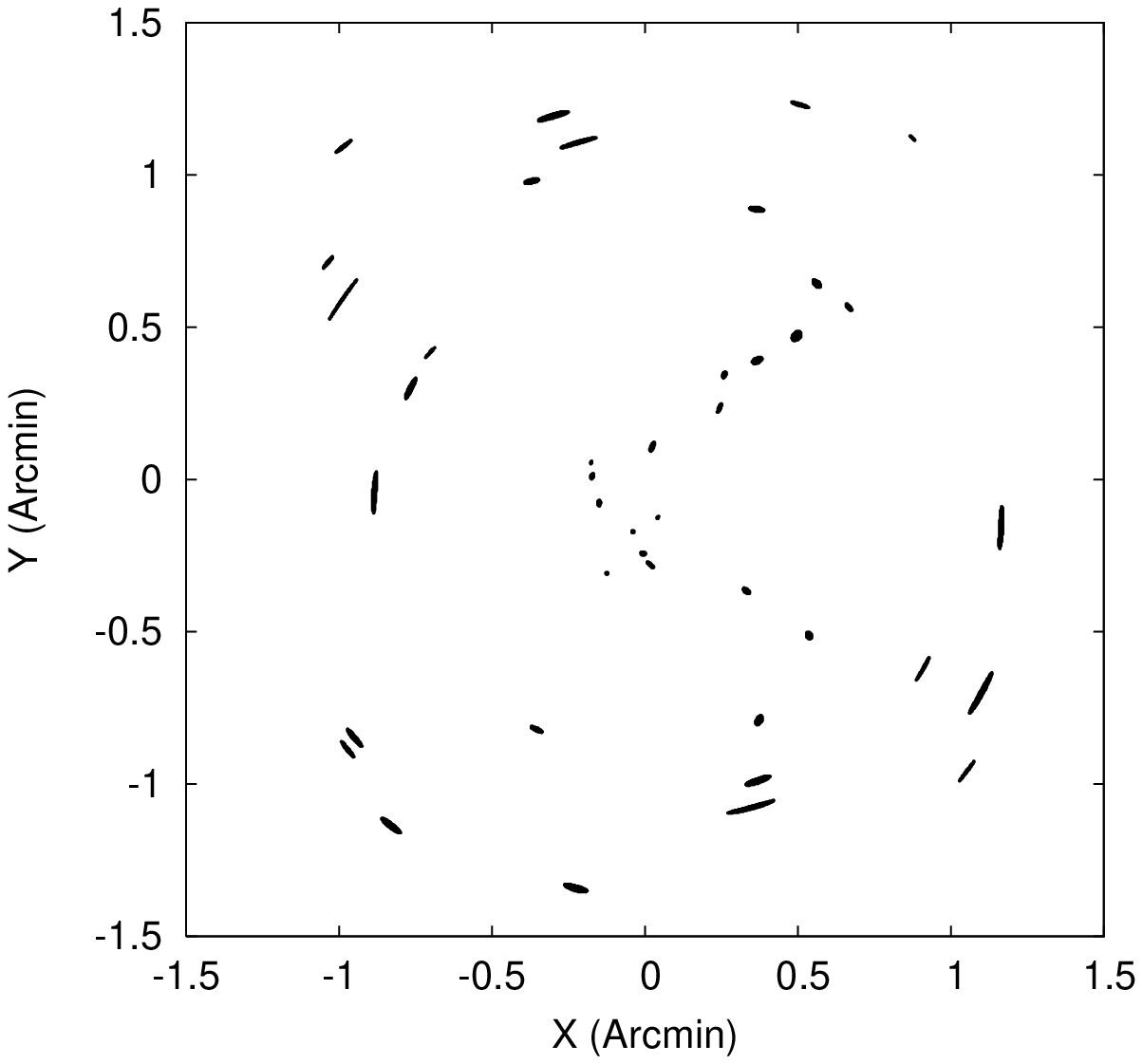}}
\subfigure{\includegraphics[width=0.48\textwidth]{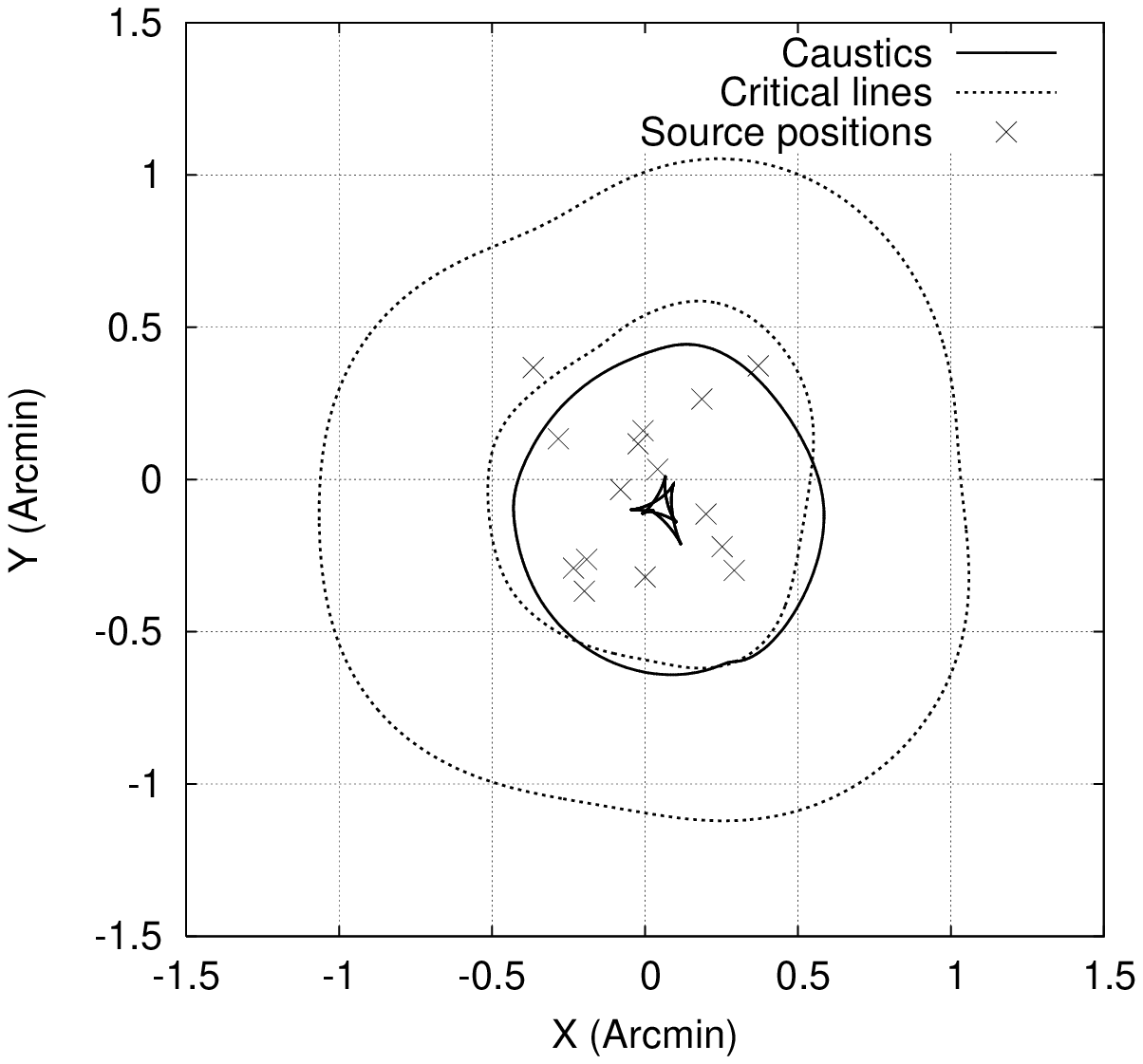}}
\addtocounter{subfigure}{-3}
\caption{Left panel:~the image used as input for the inversion
routine in the simulation. The resolution of this image is $1024 \times
1024$ pixels. Right panel:~the critical lines (dotted lines) and
caustics (full lines) of the lens for a source at redshift $z=2.5$.
The crosses indicate the positions of the sources.
}
\label{fig5}
\end{figure*}

In Fig. \ref{fig4}, we show the mass distribution and the positions
and shapes of the sources. The total mass of the lens within a radius
of $1.5${\arcmin}, which is slightly further out than the position of
the outermost image, is $0.95\times 10^{15} \,M_\odot$ and the number
of sources in this simulation is $15$. This configuration was used to
generate the images shown in the left panel of Fig. \ref{fig5}, which
in turn serves as input for the inversion algorithm. The resolution of
this image is $1024 \times 1024$ pixels. Critical lines and caustics
for a source at redshift $z=2.5$ are presented in the right panel of
Fig. \ref{fig5}, in which the source positions are also indicated. The
genetic algorithm then constructs a lens solution that projects images
of a single source onto overlapping regions in the source plane. For
this particular simulation, the fitness converged for a grid
containing about 400 Plummer mass distributions. As explained before,
multiple applications of the inversion algorithm yield different
solutions. Still, each solution manages to produce overlapping
back-projected images and, while this is in no way enforced by the
algorithm, the positions of the back-projected images are very close
to the true source positions.
\begin{figure*}
\subfigure{\includegraphics[origin=c,angle=-90,width=0.48\textwidth]{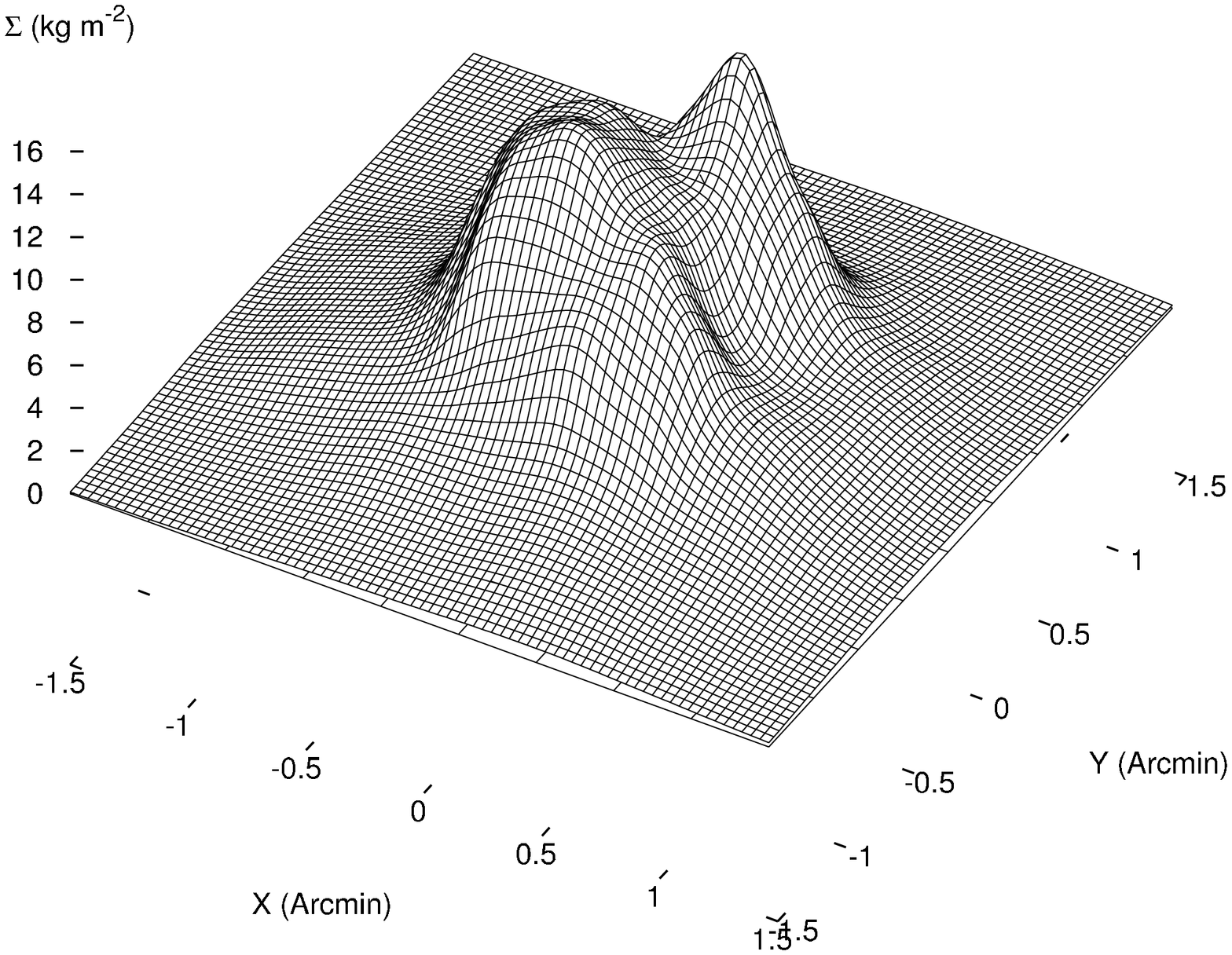}}
\subfigure{\includegraphics[width=0.48\textwidth]{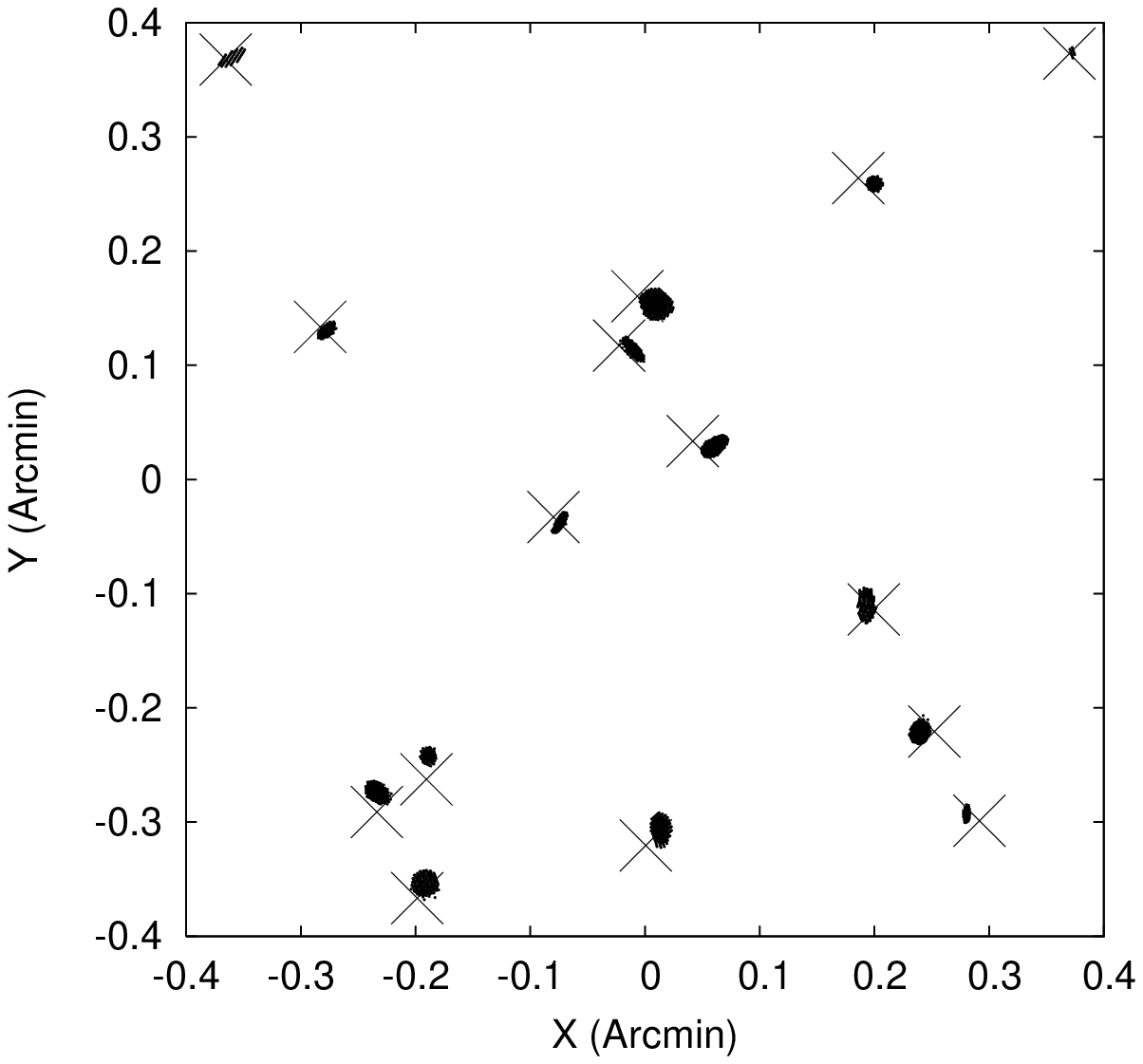}}
\caption{Left panel:~the average of 25 individual solutions for
the simulation. The total mass of this averaged lens solution within a
radius of $1.5${\arcmin} is $0.96\times10^{15}\, M_\odot$. This figure
can be compared with the left panel of Fig. \ref{fig4}. Right
panel:~the positions of the back-projected images within the source
plane for the averaged solution of the simulation. The true source
positions are marked with crosses. This figure can be compared with
the right panel of Fig. \ref{fig4}. }
\label{fig6} \end{figure*}	
\begin{figure*}
\subfigure{\includegraphics[width=0.48\textwidth]{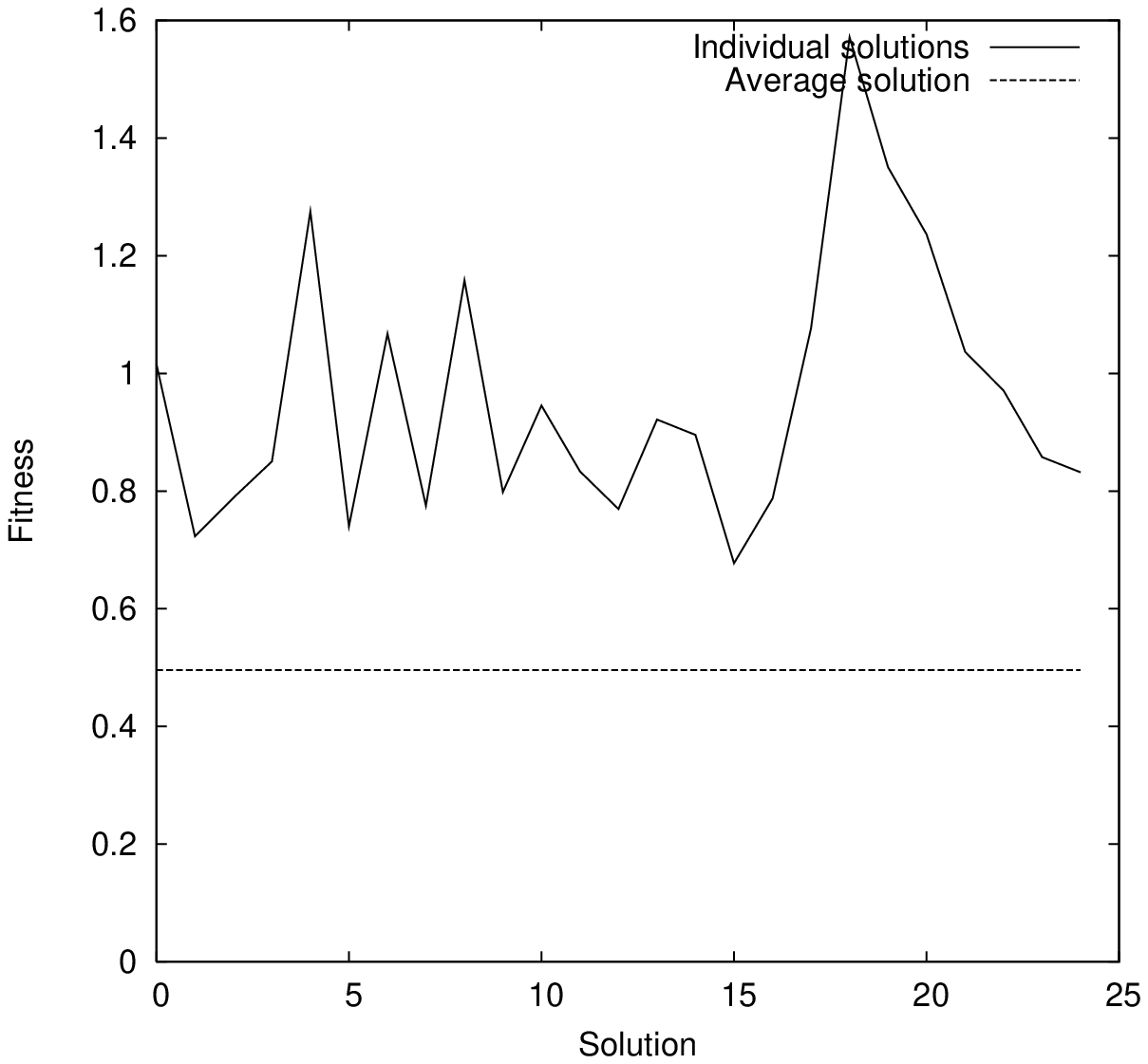}}
\subfigure{\includegraphics[width=0.48\textwidth]{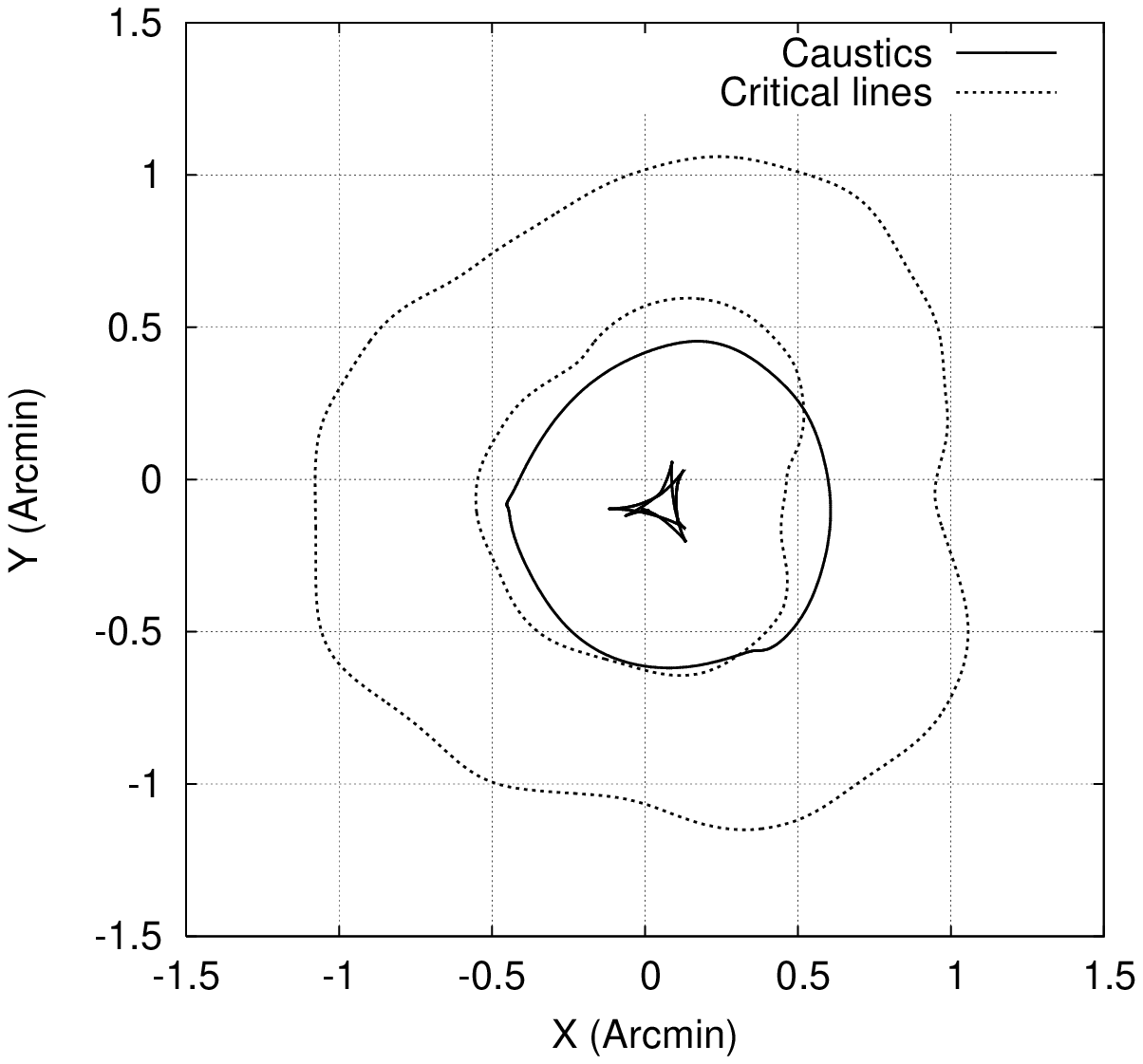}}
\caption{Left panel:~the fitness values of the 25 individual
solutions compared with the fitness value of the averaged solution for
the simulation. The averaged solution is clearly superior to the
individual reconstructions. Right panel:~the critical lines and
caustics for a source at redshift $z=2.5$ of the averaged solution in
the simulation. This figure can be compared with the right panel of
Fig. \ref{fig5}.} \label{fig7}
\end{figure*}

\begin{figure*}
\subfigure{\includegraphics[angle=-90,width=0.48\textwidth]{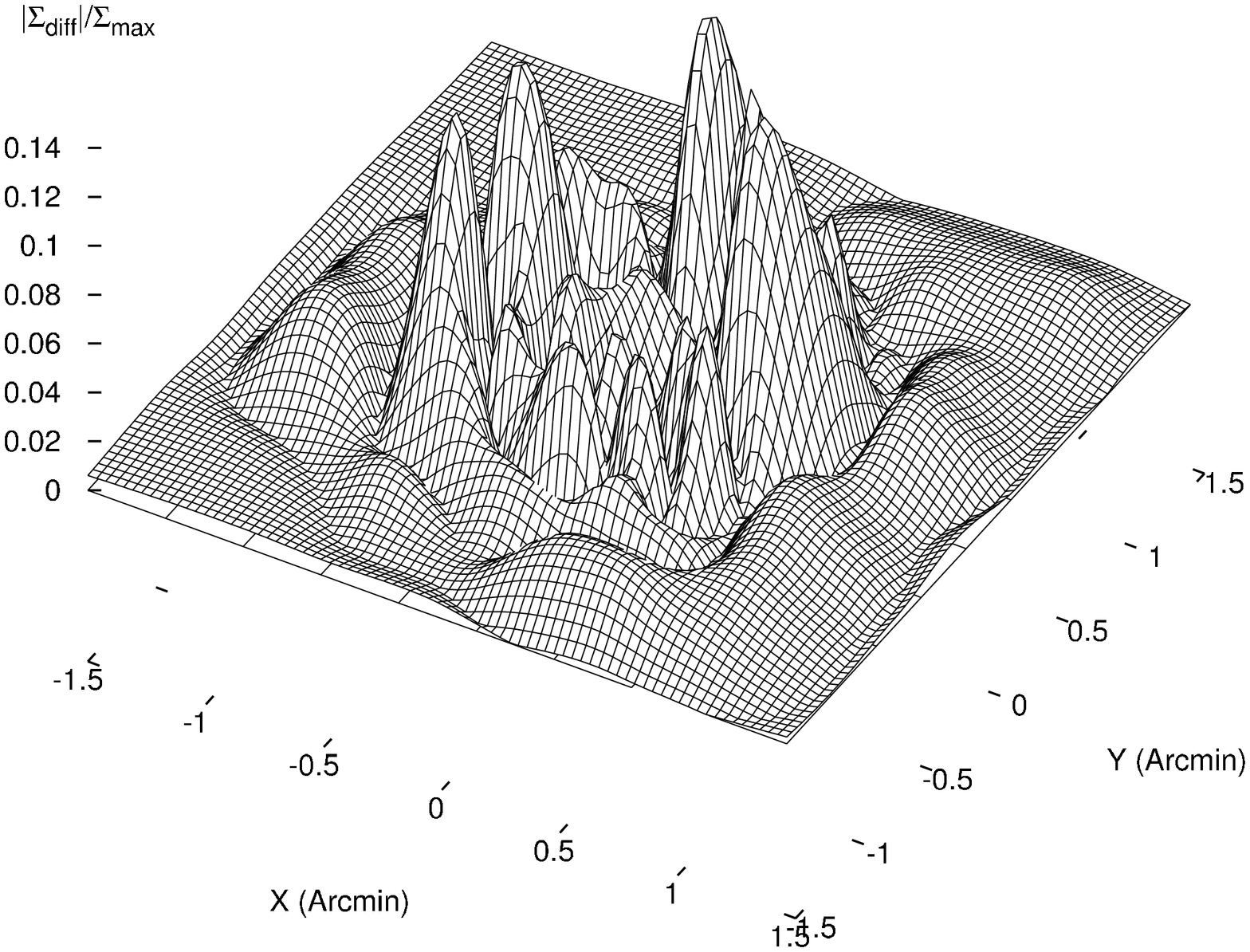}}
\subfigure{\includegraphics[angle=-90,width=0.48\textwidth]{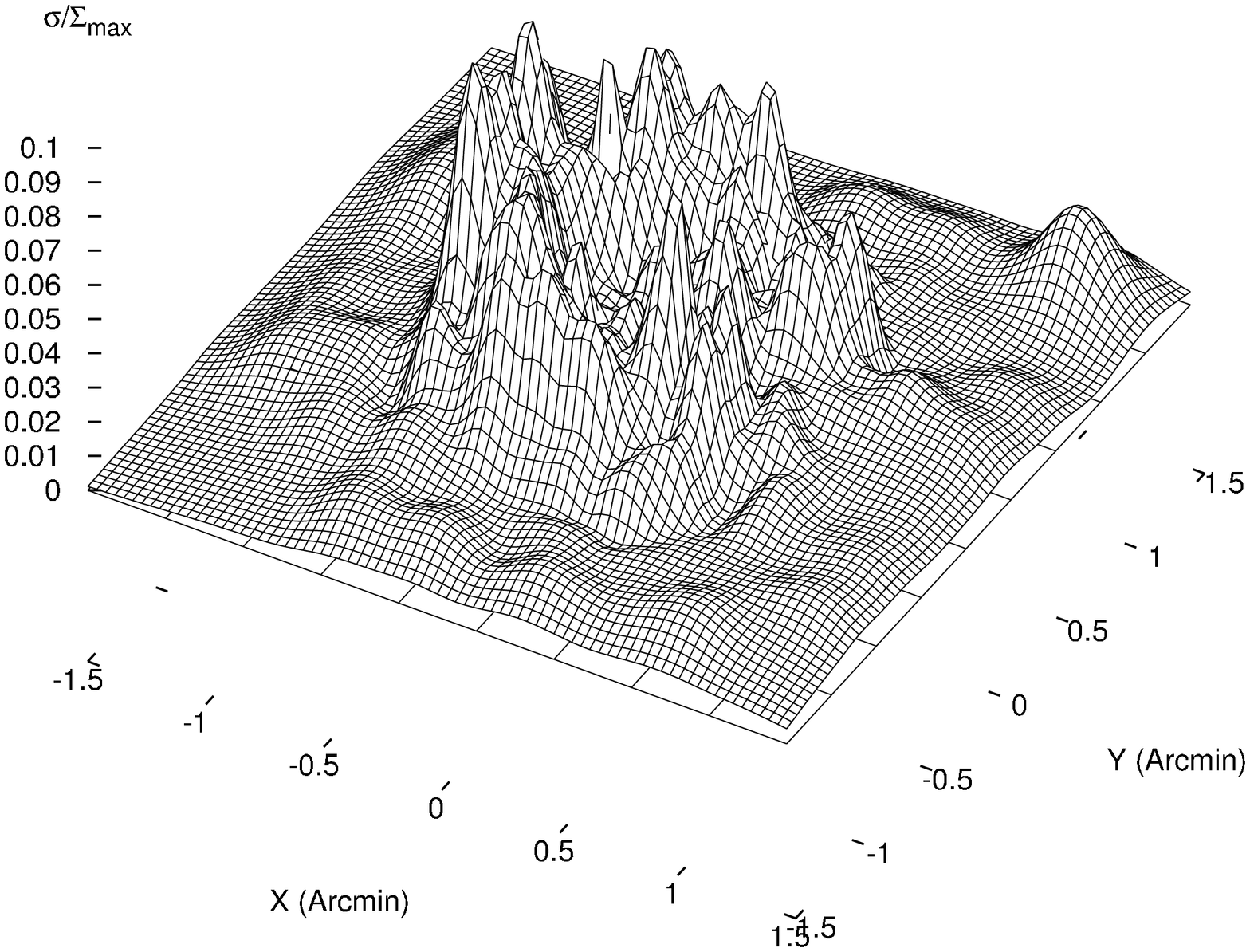}}
\caption{Left panel:~the absolute value of the difference between the
mass distributions of the input lens and of the averaged solution of
the simulation, relative to the maximum mass density of the input
lens. Right panel:~the standard deviation of the 25 individual
solutions of the simulation, relative to the maximum mass density of the
input lens. }  \label{fig8}
\end{figure*}

After applying the inversion routine $25$ times and averaging the
individual solutions, we obtained the final solution presented in the
left panel of Fig. \ref{fig6}. This figure shows a striking
resemblance to the left panel of Fig. \ref{fig4}. Clearly,
the mass distribution of the lens is retrieved with very high
accuracy. The fitness values of the 25 individual solutions and of the
averaged solution are shown in the left panel of
Fig. \ref{fig7}. Since averaging the individual solutions
suppresses random generation-to-generation fluctuations, which can
even prevent the solutions from further lowering the fitness value,
and enhances their common traits, the averaged solution outperforms
each individual solution. When the images of Fig. \ref{fig5}
are projected back onto the source plane, we obtain the situation
shown in the right panel of Fig. \ref{fig6}. The
back-projected images overlap very well and are close to the true
source positions. The critical lines and caustics of the averaged
solution for a source at redshift $z=2.5$ are shown in the right panel
of Fig. \ref{fig7}, which can be compared with the right
panel of Fig. \ref{fig5}. Again, the resemblance is
striking. In the left panel of Fig. \ref{fig8}, we show the
absolute value of the difference between the mass distributions of the
input lens and of the averaged solution. In the right panel of
Fig. \ref{fig8}, the standard deviation of the 25 individual
solutions is presented. The first quantity is a measure for the
quality of the fit, the second measures the disagreement between the
individual solutions.
	
\begin{figure}
\centering
\includegraphics[width=0.48\textwidth]{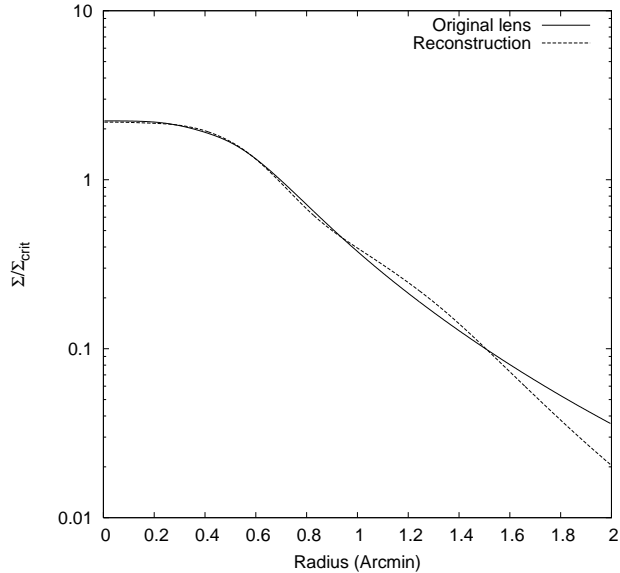}
\caption{The circularly averaged density profiles of the input lens
(full line) and of the averaged solution (dotted line), normalized to
the critical density for a source at redshift $z=2.5$. Within a radius
of 1.5{\arcmin}, which coincides with the position of the outermost
image, both profiles agree very well. Outside this radius, the density
of the averaged solution is not constrained very well by the data and
drops below that of the input lens.
\label{fig9}}
\end{figure}

For a circularly symmetric lens, only the total mass enclosed within
the radius of the outermost image can be determined. The lens employed
in the simulation is not spherically symmetric but one can still
surmise that we do not have a very good handle on the mass outside the
outermost image. In Fig. \ref{fig9}, we show the circularly averaged
density profiles of the input lens and of the averaged solution. As
expected, both agree excellently with each other within the inner
$\sim 1.5${\arcmin}, which is about the position of the outermost
image. Outside that radius, the density is no longer well constrained
by the data and the profile of the best lens solution drops below that
of the input lens. For the averaged solution, the mass enclosed within
a radius of 1.5{\arcmin} is $0.96 \times 10^{15}\,M_\odot$ which can
be compared with the input lens, which comprises a mass of $0.95
\times 10^{15}\,M_\odot$ within the same radius. 
			
From this simulation and the many others we have ran, we conclude that
given enough observational constraints, our method succeeds in
inferring the mass distribution of a lens given the redshifts of the
lens and the sources and the positions and shapes of the images. If
the sources sample the caustics sufficiently well, the density profile
of the lens can be reconstructed with great accuracy out to the
outermost image. A very important feature of our method is that it
does not minimise residuals of the lens equation, like e.g. {\sc
slap}. Because tangential images are larger than radial ones, this
residual is dominated by the tangential arcs and, as a consequence,
methods that make use of it are less sensitive to the information
contained in the radial images. This is clear from e.g. \citet{d05b},
where the non-parametrically reconstructed density profile of the
cluster A1689 shows a central decline which the authors contribute to
this effect. Our method is insensitive to the size of the images and
thus makes full use of the information contained in the radial
images. This is reflected in Fig. \ref{fig9}, which shows that the
central density of the input lens is very accurately retrieved.

\section{Discussion and conclusion}\label{sec:conc}

The procedure described and illustrated above is a non-parametric
method for inverting gravitational lenses, making no a priori
assumptions regarding the shape of the lens. We only impose the
condition that an acceptable solution must be able to map images of
the same source onto overlapping regions in the source plane. The
procedure only requires that one can identify which images correspond
to the same source. In particular, no information about the sizes of
the sources needs to be provided. The size of the grid on which the
algorithm will determine the mass distribution of the lens needs to be
specified. However, because a single solution can be obtained
relatively fast, it is an easy task to try a variety of sizes until
one is obtained which generates a lens with a good fitness value. A
multi-resolution grid with a few hundred cells is usually sufficient
to represent any plausible lens mass distribution. The implementation
used for the simulation in this paper employs a population of $250$
genomes. Based on a suite of simulations, this value proved to be
sufficiently large to to yield good solutions within an acceptable
amount of time. The calculations were done in a distibuted manner,
using sixteen Intel \textregistered{} Xeon \texttrademark{} 2.4 GHz
processors of a computer cluster. Depending on the number of sources,
creating a single solution may require several hours. To give a
specific example, the $25$ solutions used in the simulation were
created in four days.

The simulation discussed in this article, together with the many
others we performed, indicate that our inversion technique
successfully solves the lens inversion problem. The reconstructed
sources lie close to their true positions and their shapes are
retrieved quite accurately as well. The procedure determines the mass
of the lens very accurately. The averaging procedure guarantees the
removal of random fluctuations and yields a smooth mass density very
close to the true solution. Note that because the masses of the
individual Plummer distributions are always represented by positive
numbers in the genome, no negative mass densities will be produced.

Of course, the quality of the reconstruction depends on the quality of
the information at hand. The algorithm depends on the availability of
multiply imaged sources, which identifies the relevant area in our
procedure as the area within the outermost caustic. When this area is
sampled well by the sources, we can expect a good reconstruction of
the mass density, as indicated by the simulation described
previously. A major advantage of the algorithm is that it makes full
use of the information contained in the radial images, unlike methods
that minimise the residuals of the lens equation, and is thus able to
accurately reconstruct also the inner parts of the lens.

Another advantage of a genetic algorithm is the ease with which the
fitness criterion can be specified. One simply has to devise a way to
associate a fitness value with a specific genome, without worrying
about features like continuity or differentiability. E.g., to test
whether information about the amplifying effect of the gravitational
lens can improve the lens reconstruction, we added a contribution to
the fitness value:~the maximum brightness values in the back-projected
images of a single source should lie as close as possible to each
other. Several simulations indicated that augmenting the procedure in
this fashion does not improve the final result since the
$\Vec{\beta}(\Vec{\theta})$ mapping was already very well approximated
anyway. Not requiring surface brightness information not only reduces
the computational cost, but also makes the method less sensitive to
noise in the images. Incorporating information about the shear field
at larger radii outside the gravitational lens is straightforward. For
each genome the shear components $\gamma_1(\Vec{\theta})$ and
$\gamma_2(\Vec{\theta})$ can be calculated and compared with the
observed values. The expression for the fitness can be changed so as
to penalise genomes with large differences between the measured and
the model shear. Further improvements of the genetic algorithm, such
as making the mutation amplitude depend automatically on the
convergence of the fitness, are easily implemented and are left as
future research. An application to real data will be presented in a
subsequent paper.
	
\section*{Acknowledgments}

We would like to thank Prof. Philippe Bekaert and Tom Van Laerhoven of
the Expertise Centre for Digital Media for granting us access to the
computer cluster and for taking care of the related practical
issues. We also would like to thank the anonymous referee for the
valuable remarks and suggestions. They very much improved the content
and presentation of this paper.

\bsp 
\label{lastpage}
\end{document}